\documentstyle[aps,prl,multicol,fancyheadings,graphicx]{revtex}
\begin{document}

\title{Simple fluids with complex phase behavior}

\author{G.Malescio and G.Pellicane} 

\address{
Dipartimento di Fisica, Universit\`a di Messina 
and Istituto Nazionale Fisica della Materia, 98166 Messina, Italy\\
}

\date{\today}

\maketitle

\begin{abstract}

We find that a system of particles interacting through a simple isotropic potential 
with a softened core is able to exhibit a rich phase behavior including: a 
liquid-liquid phase transition in the supercooled phase, as has been suggested for water;
a gas-liquid-liquid triple point; a freezing line with anomalous reentrant behavior. 
The essential ingredient leading to these features resides in that the potential 
investigated gives origin to two effective core radii.

\end{abstract}
\medskip


\begin{multicols}{2}

Fluid-fluid or liquid-liquid transitions have been proposed for many different 
substances such as supercooled water,C,S,Ga,Se,Te,$I_2$,Cs,$SiO_2$ etc.\cite{db}. 
The transition may occur between stable phases as in carbon\cite{t,gr} or between 
metastable phases as has been suggested in supercooled water\cite{ms,be,pses}. In general,
these substances also exhibit different amorphous states as well as multiple crystalline
structures\cite{db}.
The complex intermolecular interactions characterizing the above materials are usually 
modelled through anisotropic potentials depending on the molecular orientation\cite{pot} and
thus the possibility that simple fluids interacting through isotropic potentials 
may exhibit similar exotic phase behaviors is a challenging field of investigation. 

As argued by Mishima and Stanley\cite{ms} the competition between expanded and compact 
structures in fluid-fluid (liquid-liquid) transitions suggests that the potential should
possess two equilibrium positions. The most obvious form with such feature is one with 
two wells. Such potentials were shown to give rise to water-like thermodynamic 
anomalies, though the presence of a new critical point could not be directly observed\cite{ssbs}.
Another form of interparticle interaction which could produce different equilibrium 
positions is that in which there is a region of negative curvature in the repulsive 
core: these so-called softened-core potentials were proposed by Stell and Hemmer\cite{sh}
who argued that they might produce a second transition if a first already exists.
Recently, through a mixed numerical-mean field type calculation, it was found that a 
potential consisting of a softened-core plus an infinite range van der Waals attractive
term may give rise to a second critical point\cite{j1}. 
Very recently, molecular dynamics simulation showed for a softened-core potential with 
an attractive square well evidences of a transition between two fluid phases in the 
supercooled region\cite{fmsbs}.

The purpose of this Letter is to report the findings of a study of the phase behavior 
of a system of particles interacting through a potential consisting of a hard core with
a finite repulsive shoulder and an attractive square well. Our analysis, based mainly on 
thermodynamically self-consistent ($TSC$) integral equations for fluids\cite{hm} and partly on 
Monte-Carlo ($MC$) simulations, shows the existence of a liquid-gas critical point in 
the stable fluid phase and of a liquid-liquid critical point in the supercooled region.
The liquid-gas and liquid-liquid coexistence lines meet in a gas-liquid-liquid triple 
point. Moreover, the behavior of the freezing line, estimated through one-phase 
criteria\cite{note1}, such as the Hansen-Verlet ($HV$) rule\cite{hv} and the entropic 
criterion based on the analysis of residual multiparticle entropy\cite{gg}, is 
consistent with the existence of multiple crystalline structures in the solid phase. 
This is the first time that a microscopic theory linking {\it directly}\cite{note2} the
behavior of the system to the form of the interparticle pair potential, predicts for a 
simple fluid the existence of a liquid-liquid critical point and of a gas-liquid-liquid
triple point.

The chosen potential has a repulsive part $V_{rep}(r)$ consisting of a hard core of 
radius $r_{0}= \sigma$ and a repulsive square shoulder of height $\epsilon$ and radius 
$r_{1}=2.5 \sigma$, plus an attractive component $V_{attr}(r)$ having the form of a square
well of depth $1.25 \epsilon$ extending from $r_{1}=2.5 \sigma$ to $r_{2}=3 \sigma$. 
To reach a thorough comprehension of the role played by the different components of the 
potential we first study its purely repulsive part, and then consider the effect of adding 
the attractive component.

As a preliminary investigation we study through $MC$ simulation\cite{noteMC} a system 
of particles interacting through the potential $V_{rep}(r)$ in $2D$, since in this case the 
spatial arrangement of the particles can be easily visualized. Snapshots representing the 
configuration of the system at a low temperature and different densities are shown in Fig.1, 
together with the corresponding structure factors $S(k)$.
As the density increases, the system 
rapidly undergoes a very strong ordering (Fig.1(b)) indicated by the huge first peak of 
$S(k)$ (the splitted second peak of $S(k)$ can be related to the presence of lattice 
dislocations evident in the figure). Upon further increasing the density, particles begin to 
enter the soft core and many hard cores get in contact: the first peak of $S(k)$ 
decreases while the second peak of $S(k)$ is considerably enhanced (Fig.1(d)). Finally 
at very high densities (not shown in the figure) the third peak of $S(k)$ becomes the 
highest one, while the first two peaks are greatly depressed.

The progressive ``rising and falling'' of the peaks of the structure factors reflects 
the turning on/off of different effective length scales. When the temperature $T$ and 
the density $\rho$ are sufficiently small, the soft core is practically impenetrable 
and the particles behave as hard spheres of radius $r_{1}$. As $T$ and $\rho$ increase, 
more and more particles penetrate the soft core until this becomes scarcely influent and 
the system is essentially equivalent to an assembly of hard spheres of radius $r_{0}$. 
In general, the system can be considered a ``mixture'' of two populations of hard 
spheres, of radius $r_{0}$ and $r_{1}$ respectively. The relative concentration of the 
two species is fixed by the values of $T$ and $\rho$.
Thus, in contrast to standard simple fluids, the system has three possible 
length scales: $r_{1}$, $r_{10}=(r_{1}+r_{0})/2$ and $r_{0}$, and as many indicators of 
structural ordering, namely the peaks of the structure factor corresponding to the 
wavevectors $k_1$, $k_{10}$, $k_0$, associated to these lengths. 
We stress the crucial role of the finite repulsive shoulder; due to its presence 
particles can be in one of two ``states'': this is the essential feature which opens 
the possibility of liquid-liquid-immiscibility in a pure substance.

Let us now consider a system of particles interacting through the potential $V_{rep}(r)$ 
in $3D$. We study its structural and thermodynamical properties using the $TSC$ 
Roger-Young ($RY$) integral equation\cite{ry}. According to $HV$ rule a fluid is 
expected to undergo crystallization when the first (main) peak of $S(k)$ attains the 
value $2.85$\cite{hv}. This statement refers to simple fluids with a single length scale. 
In our case, different length scales come into play and one must consider all the 
associated indicators of structural ordering. As shown in Fig.2, their behavior, upon 
increasing the density at constant $T$, appears analogous to that observed in $2D$, with each peak 
in turn rising and then ``falling'' down (except that associated to the hard core radius
which becomes higher and higher as $\rho$ increases). The overall trend is confirmed by
$MC$ calculation, though the theory\cite{noteint} tends to overestimate the heights of 
$S(k_{10})$ and $S(k_0)$ at high densities.

In Fig.3 we show the locii of the points of the plane $T$,$\rho$ for which $S(k_1)$, 
$S(k_{10})$ and $S(k_0)$ are equal to $2.85$. The freezing line predicted through a 
straightforward application of $HV$ rule to the ``anomalous'' simple fluid investigated,
should coincide with the line which bounds the region where at least one of the peaks 
of the structure factor is greater than $2.85$. This line shows a reentrant behavior 
which suggests that the system may undergo crystallization upon the density {\it decreasing}. 
The behavior can be related to the presence of the finite repulsive shoulder. In fact, the 
fraction of particles penetrating the soft core increases with the density (at constant
$T$) thus ``generating'' additional space for the system. This effect is particularly 
important where one length scale begins to become less effective in favor of the smaller
one. In these regions the phenomenon may overcompensate the general decrease of the 
space available to the system upon the density increasing thus causing a tendency of the 
system to become less ordered: accordingly, the freezing line may have a negative 
derivative. We note, however, that this behavior might be overestimated in Fig.3.
In fact, since different length scales are contemporarily effective and contribute to 
the global structural ordering of the system, we expect that a more realistic freezing line 
has a smoother behavior which ``blends'' together the upper portions of the different 
curves of Fig.3.

The inset in Fig.3 shows a magnification of the low temperature-low density region.
Here, since only one length scale ($r_1$) is effective, the prediction of $HV$ is 
expected to be reliable. The freezing line starts nearly vertical at a density 
$\rho^*\simeq 0.06$ which corresponds to the freezing density $\rho r_{1}^3 \simeq 0.945$ 
of a fluid of hard spheres of radius $r_1$. As $\rho$ increases, the freezing line bends 
and shows a reentrant behavior: the phenomenon is associated with the very onset of the 
soft core penetration which, for the reasons above discussed, has a disordering effect on the system.
This is reflected in the anomalous behavior shown by the isothermal compressibility 
$\chi_T$, which suddenly increases with the density (see same figure).
Correspondingly, the first peak of $S(k)$ abruptly decreases: 
this occurs also at lower temperatures (where the value of $S(k_1)$ remains 
however greater than $2.85$) along a line coinciding approximately with the extension 
of the portion of the freezing line with negative derivative\cite{mp}. This line meets 
the $T=0$ axis at a density $\rho^* \simeq 0.09$ which corresponds to the closest 
packing of hard spheres of radius $r_{1}$ (occurring at $\rho r_{1}^3 = \sqrt 2$). 
The above results lead to conclude that the region shadowed in the inset of Fig.3 
corresponds to an {\it expanded} solid phase of the system.
                                  
We now investigate the phase behavior of a system of particles interacting through 
the full potential $V_{rep}(r)$ plus $V_{attr}(r)$. Calculations are performed making use 
of the $HMSA$ $TSC$ equation\cite{zh}, which is better suited than $RY$ equation for 
interparticle interactions including an attractive component and reduces to this for 
purely repulsive potentials.
The phase diagram of the system is shown in Fig.4. Two coexistence curves occur\cite{note3}, each 
terminating at a critical point, denoted $C1$ and $C2$. The critical densities and 
temperatures are respectively $\rho^*_{C1}=0.06$, $T^*_{C1}=1.3$, and $\rho^*_{C2}=0.77$, 
$T^*_{C2}=0.55$. These values were estimated using the rectilinear diameter rule and the 
scaling relationship for the width of the coexistence curve with the non-classical exponent
$\beta \approx 0.325$\cite{st}. Below $T_{C1}$ the system separates into a gas and a liquid phase. 
The liquid phase is not unique since, below $T_{C2}$ ($T_{C2}<T_{C1})$, separates into distinct 
low-density ($LD$) and high-density ($HD$) phases. 
Since the critical point $C2$ is well below the freezing line, the 
liquid-liquid transition occurs between metastable phases in the supercooled region of the system. 
This feature recalls the scenario proposed for water\cite{ms,pstsa}, but in that case the 
liquid-liquid coexistence line is expected to start from $C2$ running at higher pressures as $T$ 
decreases\cite{pses}. In the system investigated, the contrary is observed, this line running at lower 
pressures as $T$ decreases (see inset of Fig.4). This makes possible a new feature:
the simultaneous coexistence of three fluid phases. In fact, the gas-liquid and the 
liquid-liquid coexistence lines meet in a gas-liquid-liquid triple ($GLL$) point which lays in 
the supercooled phase ($\rho^{*}_{GLL}\approx 0.57$, $T^{*}_{GLL}\approx 0.53$).

The addition of the attractive well causes a shift towards higher temperatures of the locii of 
the points where $S(k_1)$, $S(k_{10})$ and $S(k_0)$ are equal to $2.85$, while their location 
in density remains essentially unaltered. Only a small portion of the line $S(k_1)=2.85$ is 
visible since it lays almost entirely in a region, correspondingly approximately to the 
gas-liquid spinodal decomposition, where the theory is unstable. 
The freezing line corresponding to $HV$ rule and that estimated through the entropic 
criterion are in substantial agreement at low and high densities (pressures),
yielding close estimates of the gas-liquid-solid ($GLS$) triple point
($\rho^{*}_{GLS}\approx 0.29$, $T^{*}_{GLS}\approx 0.97$). On the contrary, these lines 
differ markedely in the intermediate region, where the first one shows a very evident 
reentrant behavior, while in the second approach this can be appreciated only numerically\cite{mp}. 
For the reasons exposed in the purely repulsive case, the second result is probably more 
reliable. Thus, as shown in the inset in Fig.4, the freezing temperature increases initially with
pressure, then remains essentially constant (according to the entropic criterion) in the pressure 
range $1 \leq P^{*} \leq 2$, ($P*=P \sigma^3/\epsilon$) and eventually increases again with pressure.
Though our estimate of the freezing line is based solely on one-phase criteria, its shape, with 
branches having distinctly different slopes, is consistent with the possibility that structural changes
occur in the solid state of the system. Consequently, transitions may be possible between solid 
phases of the system investigated.

The results presented in this Letter show that a pure model system, with a softened-core
isotropic potential, may have a rich phase behavior with features typical of substances
characterized by much more complex anisotropic interactions. 
The possibility that simple substances related to the model investigated 
may exist (or may be ``realized'') in nature well deserves to be investigated.

\smallskip

We wish to thank
F. Sciortino, 
P.Ballone,
P.V. Giaquinta, 
C.Caccamo,
S.Dugdale,
for helpful suggestions and for interesting and stimulating discussions. 
One of us (G.M.) wishes to thank H.E.Stanley for stimulating his interest on the subject.

\end{multicols}

\newpage

\begin{figure}
\caption{$2D$ system. Snapshots showing the configurations calculated through
$MC$ simulation at a reduced temperature $T^*=k_B T/\epsilon = 0.05$  
($k_B$ is the Boltzmann constant) and different reduced densities
$\rho^*=\rho \sigma^D$ ($D$ is the dimensionality of the system and $\rho$ 
is the number density of the particles). 
Below each snapshot the corresponding structure factor is shown.
Dots represent the hard core of the particles. }     
\end{figure}
\label{fig:1}

\begin{figure}
\begin{center}
\includegraphics[width=18cm,angle=90]{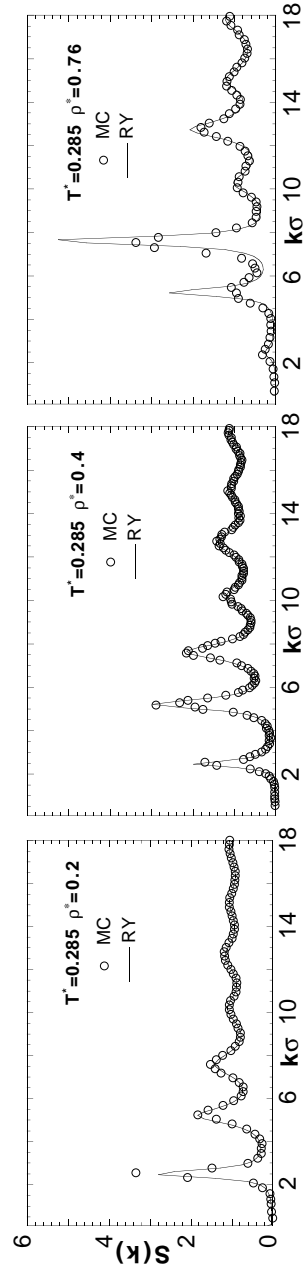}
\end{center}
\caption{Purely repulsive potential. Structure factors within the $RY$ equation and 
$MC$ simulation.}              
\end{figure}
\label{fig:2}

\begin{figure}
\begin{center}
\includegraphics[width=18cm,angle=90]{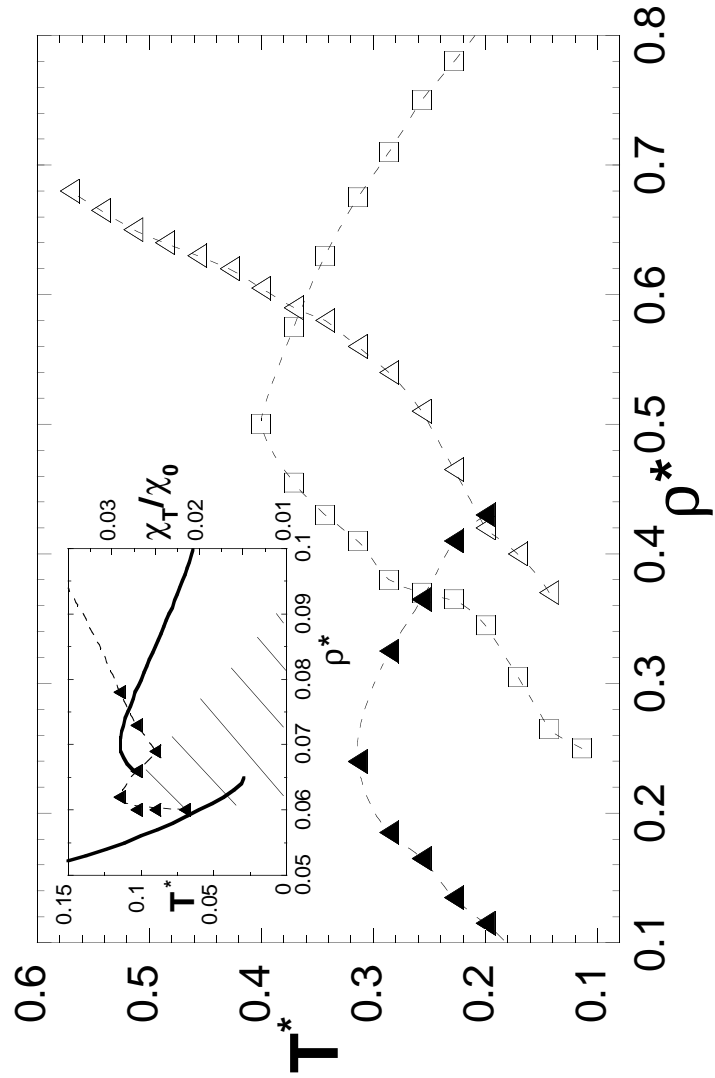}
\end{center}
\caption{Purely repulsive potential. The dashed lines are the locii of points of the $T$,$\rho$
plane where $S(k_1)$ (full triangles), $S(k_{10})$ (squares) and $S(k_0)$ (open triangles) are
equal to $2.85$. 
The inset shows a magnification of the low $T$-low $\rho$ region; the isothermal 
compressibility $\chi_T$ (thick solid line) is calculated along the isotherm $T^*=0.103$ 
($\chi_0$ is the ideal gas compressibility).}
\label{fig:3}
\end{figure}

\begin{figure}
\begin{center}
\includegraphics[width=18cm,angle=90]{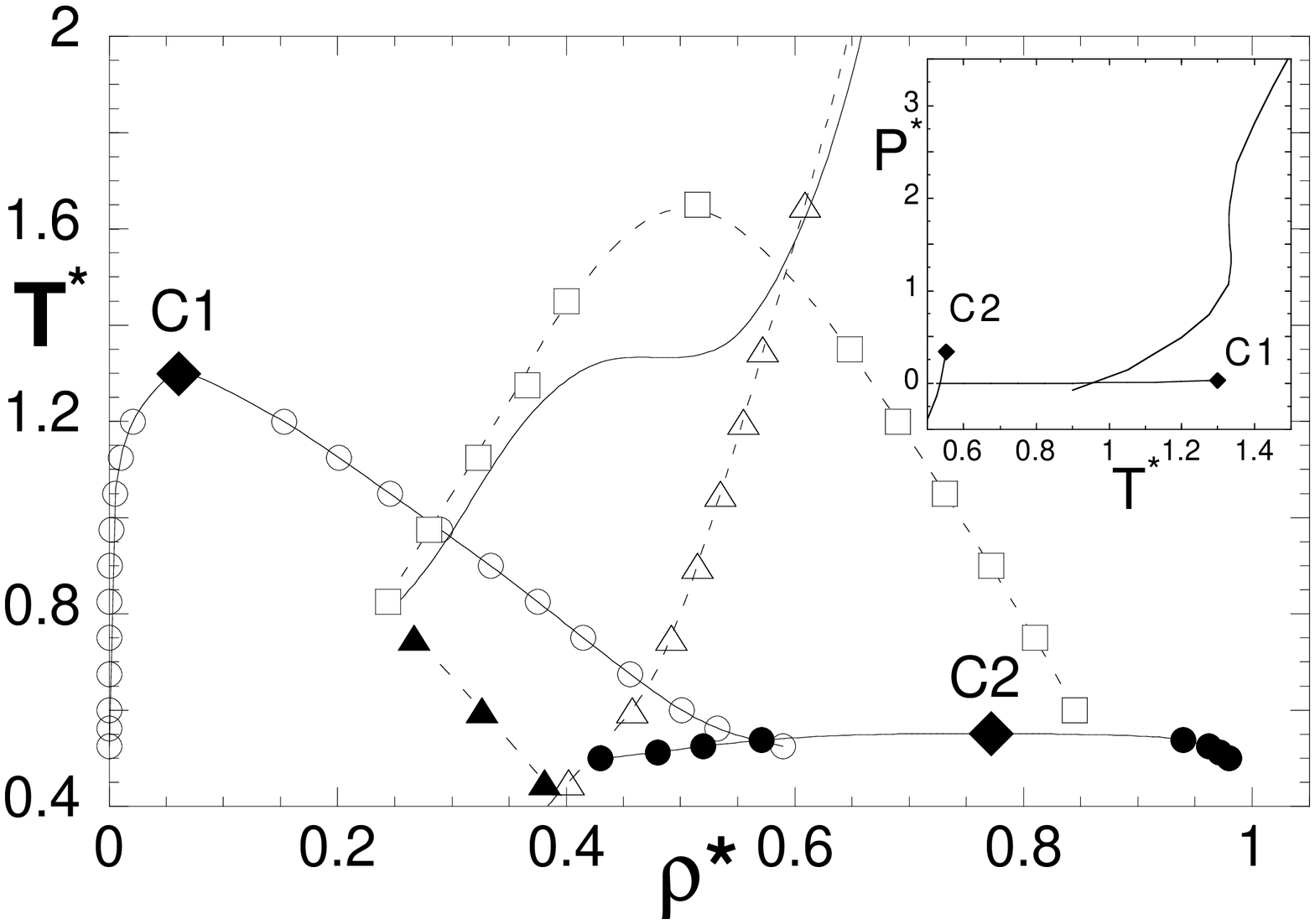}
\end{center}
\caption{Phase diagram in the $T$,$\rho$ plane. Coexistence lines: gas-liquid (circles), 
liquid-liquid (dots). The full diamonds represent the critical points.
The dashed lines are the locii of points where 
$S(k_1)$ (full triangles), $S(k_{10})$ (squares) and $S(k_0)$ (open triangles) are equal to 
$2.85$. The solid line with no symbols is the freezing line estimated through the entropic 
criterion. Inset: $P$,$T$ phase diagram; pressure is given in units of $\epsilon/\sigma^3$.}
\label{fig:4}
\end{figure}


\smallskip 

\begin{thebibliography}{99}


\bibitem{db}
P.G.Debenedetti, {\it Metastable liquids: concepts and principles}, 
Princeton Univ. Press, Princeton (1998);
K.Tsuji, J.Noncryst. Solids {\bf 117-118}, 27 (1990);
K.Yaoita et al., {\it ibid.} {\bf 156-158}, 157 (1993); {\bf 150}, 25 (1992);
V.V.Brazhkin et al., High Press. Res. {\bf 6}, 363 (1992);
Phys.Lett.A {\bf 154}, 413 (1991).

\bibitem{t}
M.Togaya, Phys.Rev.Lett. {\bf 79}, 2474 (1997).

\bibitem{gr}
J.N.Glosli and F.H.Ree, Phys.Rev.Lett. {\bf 82}, 4659 (1999)

\bibitem{ms} 
O.Mishima and H.E.Stanley, Nature {\bf 396}, 329 (1998);
ibid {\bf 392}, 164 (1998).

\bibitem{be}
M.C.Bellisent-Funel, Europhys.Lett. {\bf 42}, 161 (1998).

\bibitem{pses}
P.H.Poole, F.Sciortino, U.Essmann and H.E.Stanley, Nature
{\bf 360}, 324 (1992). 

\bibitem{pot}
F.H.Stillinger and A.Rahman, J.Chem.Phys. {\bf 60}, 1545 (1974);
W.L.Jorgensen,J.Chandrasekhar,J.Madura,R.W.Impey and M.Klein,
J.Chem.Phys. {\bf 79}, 926 (1983);
D.W.Brenner, Phys.Rev.B {\bf 42}, 9458 (1990); {\bf 46}, 1948 (1992).

\bibitem{ssbs}
M.R.Sadr-Lahijany, A.Scala, S.Buldyrev and H.E.Stanley, 
Phys. Rev. Lett. {\bf 81}, 4895 (1998);

\bibitem{sh} 
P.C.Hemmer and G.Stell, Phys. Rev. Lett. {\bf 24}, 1284 (1970).

\bibitem{j1} 
E.A.Jagla, J. Chem. Phys. {\bf 101}, 8980 (1999).

\bibitem{hm} 
J.P.Hansen and I.R.McDonald {\it Theory of simple liquids} (Academic Press, London) 
(1976).

\bibitem{fmsbs}
G.Franzese,G.Malescio,A.Skibinsky,S.V.Buldyrev and H.E.Stanley,
(cond-mat 0005184).

\bibitem{note1}
The use of one-phase criteria is motivated by the fact that, due to the softened core, 
there are many possible solid phases and in general it is not easy to tell safely which
is the most stable structure\cite{j2}, except in the limit case of a very narrow shoulder
which can be treated perturbatively\cite{bf}. 

\bibitem{j2}
E.A.Jagla, J.Chem.Phys. {\bf 110}, 451 (1999).

\bibitem{bf}
P.Bolhuis and D.Frenkel, J.Phys.Condens.Matter {\bf 9}, 381 (1997).

\bibitem{hv}
J.P.Hansen and L.Verlet, Phys.Rev. {\bf 184}, 151 (1969).

\bibitem{gg}
P.V.Giaquinta and G.Giunta, Physica A {\bf 187}, 145 (1992).

\bibitem{note2}
Preceding theoretical studies showing phase diagrams with a second critical point were based 
on extensions of the van der Waals theory which assumed some appropriate functional form of 
the free energy\cite{pstsa} or derived it from effective density dependent potentials\cite{tb}. 

\bibitem{pstsa}
P.H.Poole, F.Sciortino, T.Grande, H.E.Stanley and C.A.Angell, 
Phys. Rev. Lett. {\bf 73}, 1632 (1994).

\bibitem{tb}
C.F.Tejero and M.Baus, Phys. Rev. E {\bf 57}, 4821 (1998).

\bibitem{noteMC}
$MC$ calculations were performed using a standard Metropolis algorithm\cite{at} within
the $NVT$ ensemble for a system of $1000$ particles. For each thermodynamic state 
equilibrature cycles consist of at least $10^6$ $MC$ steps in $2D$ and $2 \times 10^5$ 
$MC$ steps in $3D$. A single $MC$ step consists in an attempt to move all the particles 
with acceptance rate between $0.3$ and $0.4$. The structure factor is 
calculated through cumulation runs of $2 \times 10^4$-$10^5$ $MC$ steps (depending on the density) 
through the relationship $S(k)=<\rho(k) \rho(-k)>$. 
Test runs with $5000$ particles were performed to check for finite size effects.

\bibitem{at}
M.P.Allen and D.J.Tildesley, {\it Computer simulation of liquids},
Oxford Univ. Press, London (1987).

\bibitem{ry}
F.J.Rogers and D.A.Young, Phys.Rev.A {\bf 30}, 999 (1984).

\bibitem{noteint}
Solutions of integral equations were obtained through Gillan's algorithm\cite{gil}
on a grid of $2048$ points spanning over more than $20$ $\sigma$ in the direct $r$-space.

\bibitem{gil}
M.J.Gillan, Mol.Phys. {\bf 38}, 1781 (1979).

\bibitem{mp}
G.Pellicane and G.Malescio, (unpublished results).

\bibitem{zh}
G.Zerah and J.P.Hansen, J.Chem.Phys. {\bf 84}, 2336 (1986).

\bibitem{note3} 
The coexistence curves are calculated by equating, at constant $T$ and $P$, the chemical
potentials of the coexisting phases.
Generally, $TSC$ integral equations are derived assuming that local consistency is 
equivalent to global consistency\cite{zh}. Since in the low $T$-high $\rho$ region this
approximation is not very satisfactory and global consistency is unpracticable, the
pressure is calculated along mixed isochore-isothermal paths which minimize the isochore 
portions where the more accurate ``compressibility route'' pressure cannot be evaluated.

\bibitem{st}
J.M.Yeomans, {\it Statistical mechanics of phase transitions},
Clarendon Press, Oxford (1992).


\end{thebibliography}
\end{document}